\documentclass[aps,prl,twocolumn,groupedaddress]{revtex4}

\usepackage{graphicx}
\usepackage{amssymb}

\newcommand{\rs}{\rm \scriptscriptstyle}

\begin{document}

\title{Signature   of Quantum Depletion in the Dynamic Structure Factor  of Atomic Gases}

\author{H.P.\ B\"uchler}
\author{G.\ Blatter}
\affiliation{Theoretische Physik, ETH-H\"onggerberg, CH-8093
Z\"urich, Switzerland}

\date{\today}

\begin{abstract}
  We study the linear response and the dynamic structure factor of
  weakly interacting Bose gases at low temperatures. Going beyond
  lowest order in the weak coupling expansion allows us to determine the
  contribution of the thermal and quantum depletion of the condensate to the
  dynamic structure factor. We find that the quantum depletion
  produces a pronounced peak in the dynamic structure factor,
  which allows for its detection via a
  spectroscopic analysis.
\end{abstract}


\maketitle

The main characteristic of bosonic quantum gases is the formation
of a condensate at low temperatures as observed in $^{4}{\rm He}$
and atomic gases \cite{griffin93,anderson95}. In superfluid
$^{4}{\rm He}$, strong interactions produce a large depletion of
the zero-momentum state and the condensate involves only a small
fraction of the particle density; the condensate was observed
experimentally by measuring the dynamic structure factor with
neutron scattering \cite{griffin93,henshaw61}. In turn,
interactions are weak in atomic gases and the condensate involves
nearly the full particle density, while the quantum depletion is
small. However, recent progress in tuning atomic gases allows to
increase the interaction strength via Feshbach resonances
\cite{cornish00} or optical lattices \cite{jaksch98,greiner02}. It
is this strongly correlated regime with large quantum depletion
which is currently attracting a lot of experimental and
theoretical interest. In this letter, we study the contribution of
the quantum depletion to the dynamic structure factor, and propose
that, in analogy to superfluid $^{4}{\rm He}$, spectroscopic
studies provide a powerful tool for characterizing the structure
of atomic gases and allow for the detection of the quantum
depletion.

\begin{figure}[hbtp]
\vspace{-0.cm}
\includegraphics[scale=0.38]{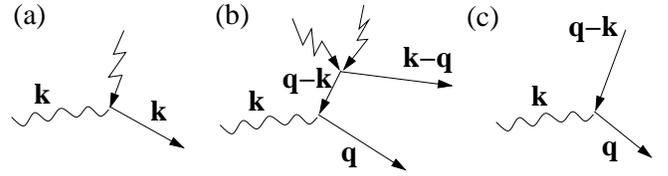}
\vspace{-0.2cm}
  \caption{ \label{dissipativeprocesses}
  Diagrams representing the excitations created by  the external
  drive $\delta V$ (wiggly line).
  (a) Creation of a quasi-particle (straight line)
  from the condensate $n_{0}$ (zigzag line). (b) Creation of two
  quasi-particles probing the quantum depletion.
  (c) Creation of a particle-hole pair probing the thermal depletion.}
\end{figure}

The dynamic structure factor of an ideal bosonic gas exhibits a
delta function peak  $S(\omega,{\bf k})= n \:\delta(\omega -
\epsilon_{{\bf k}})$, thus probing excitations from the condensate
with the energy $\epsilon_{{\bf k}}= \hbar^{2} {\bf k}^2/2m$, see
Fig.~\ref{dissipativeprocesses}(a). Previous studies of the
dynamic structure factor in atomic Bose gases accounted for the
modification of such single-particle excitations via interaction
and finite trapping: while the interaction modifies the
quasi-particle excitation spectrum and the weight of the delta
function peak \cite{stenger99.2,stamper-kurn99,steinhauer02}, the
finite size of the trap induces a broadening of this peak
\cite{wu96,zambelli00}.
In this letter, we determine the linear response function and the
dynamic structure factor of weakly interacting Bose gases within a
microscopic analysis going to first order in the interaction
parameter $\sqrt{n a^{3}}$ (here, $n$ is the particle density and
$a$ denotes the scattering length). We find, that the structure
factor contains two contributions: The first accounts for
excitations of quasi-particles from the condensate, see
Fig.~\ref{dissipativeprocesses}(a), and involves the condensate
density $n_{0}$. The second term is due to excitations of two
quasi-particles and accounts for the quantum depletion $n_{\rs
D}=1-n_{0}$ of the condensate, see
Fig.~\ref{dissipativeprocesses}(b); particle-hole excitations [see
Fig.~\ref{dissipativeprocesses}(c)] only contribute to the
structure factor at finite temperatures. The quantum depletion
$n_{\rs D}$ is well understood within the weak coupling Bogoliubov
theory and accounts for the interaction-induced expulsion of
particles from the zero momentum state. Although the extension of
this theory to finite traps is successful in explaining many
phenomena observed in cold atomic gases, the detection of the
quantum depletion $n_{\rs D}$ itself has not been achieved so far.
We show below that the quantum depletion produces a pronounced
peak in the dynamic structure factor allowing for its observation
via a spectroscopic analysis.

The microscopic Hamiltonian of interacting bosons takes the form
\begin{eqnarray}
  H &=& \int d{\bf x} \;\psi^{+}({\bf x}) \left(-
  \frac{\hbar^{2}\Delta}{2m}\right)\psi({\bf x})\\ & & + \int d{\bf
  x}d{\bf y} \; \hat{U}({\bf x}-{\bf y})\psi^{+}({\bf x})\psi^{+}({\bf y}) \psi({\bf y})
  \psi({\bf x}) \nonumber
\end{eqnarray}
with $\psi^{+}({\bf x})$ and $\psi({\bf y})$ the bosonic field
operators and the interaction potential $\hat{U}({\bf x})$. In the
low-density limit, the microscopic interaction potential
$\hat{U}({\bf x})$ is replaced by the $s$-wave scattering
potential $U({\bf x})=g \delta({\bf x})$ with $g =4 \pi \hbar^{2}
a/m$ \cite{hugenholtz59,abrikosovbook,nozieresbook2}. The
expansion parameter in this weakly interacting Bose gas is
$\sqrt{n a^{3}}$.

The response $\delta \rho(t,{\bf x})$ of the bosonic density
relates to a small external potential $\delta V(t,{\bf x})$  via
the integral relation
\begin{equation}
 \delta \rho(t,{\bf x}) = \int dt' d{\bf x}' \chi(t-t',{\bf x}-{\bf
  x}')\delta V(t',{\bf x}')
\end{equation}
with the linear response function \cite{negele98} ($\rho= \psi^{+}
\psi$)
\begin{equation}
  \chi(t,{\bf x})= - i \Theta(t)
  \langle \left[ \rho(t,{\bf x}),\rho(0,0)\right]\rangle .
\end{equation}
Here, $\langle \ldots \rangle$ denotes the quantum statistical
average at fixed temperature $T$ and chemical potential $\mu$. The
response function  $\chi(t,{\bf x})$ is conveniently calculated
via its relation to the density-density correlation function in
imaginary time $D(\tau,{\bf x})$ \cite{negele98},
\begin{eqnarray}
 D(\tau,{\bf x})&=& - \Big[\big\langle  T \rho(\tau,{\bf x})
 \rho(0,0)  \big\rangle
- \big\langle \rho \big \rangle \big\langle \rho
 \big\rangle \Big]\label{densitydensitycorrelator}
\end{eqnarray}
with $T$ the time ordering operator. Then, the response
$\chi(\omega,{\bf x})$ is the proper analytic continuation of the
density-density correlation function $D(\Omega_{s},{\bf x})$,
which respects the retarded character of $\chi(t,{\bf x})$. Here,
$\Omega_{s}=2\pi T s$ denote the Matsubara frequencies with $s\in
\mathbb{Z}$.

\begin{figure}[hbtp]
\vspace{-0.cm}
\includegraphics[scale=0.35]{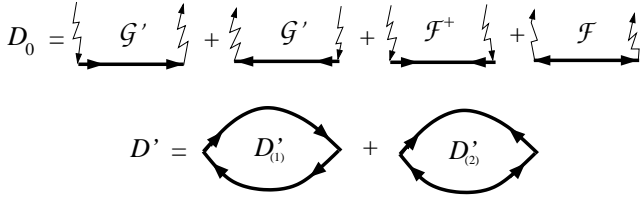}
\vspace{-0.2cm}
  \caption{ \label{Dcombineddiagram} Diagrams contributing to the
  response function $D$ to first order in $\sqrt{n a^{3}}$. $D_{0}$
  collects diagrams involving the condensate density $n_{0}$, while
  the diagrams in $D'$ probe the depletion of the condensate.
  The thick solid line denotes the normal and anomalous Green's
  functions $\mathcal{G}$, $\mathcal{F^{+}}$, and $\mathcal{F}$ within
  Bogoliubov approximation. Zigzag lines account for the creation
  $\xi^{+}$ and annihilation $\xi$ of particles from the condensate.}
\end{figure}

In the following, we apply standard quantum field theory methods
for interacting bosons at low temperatures
\cite{hugenholtz59,abrikosovbook,nozieresbook2}. We separate the
bosonic operators $\xi^{+}$ and $\xi$ accounting for creation and
annihilation of the zero momentum state from the field operator,
\begin{equation}
  \psi^{+}(\tau,{\bf x})=\xi^{+} + \psi'^{+}(\tau,{\bf x}),
  \hspace{10pt}\psi(\tau,{\bf x})=\xi + \psi'(\tau,{\bf x}).
  \label{fieldoperator}
\end{equation}
Within each diagrammatic expression the operators $\xi$ and
$\xi^{+}$ are considered as $c$-numbers contributing a factor
$\sqrt{n_{0}}$; $n_{0}$ denotes the condensate fraction and has to
be determined self-consistently. The difference $n_{\rs D}=
n-n_{0}$ between the averaged particle density $n=\langle
\rho\rangle$ and the condensate fraction $n_{0}$ is due to the
thermal and quantum depletion of the condensate. Here, we focus on
the low density limit and expand the density-density correlation
function (\ref{densitydensitycorrelator}) in the small parameter
$\sqrt{n a^{3}}$; the relevant diagrams are shown in
Fig.~\ref{Dcombineddiagram}. All additional diagrams involve a
vertex operator $\Gamma$ and contribute to higher order terms in
the expansion parameter $\sqrt{na^{3}}$; they will be dropped in
the following.

The diagrammatic representation of the leading contribution
$D_{0}$ to the density-density correlation function $D$ is shown in
Fig.~\ref{Dcombineddiagram} and takes the form
\begin{eqnarray}
   D_{0}(\Omega_{s},{\bf k})&=&- n_{0} \Big(
   \mathcal{G'}(\Omega_{s},{\bf k})+\mathcal{G'}(-\Omega_{s},-{\bf
   k})\\&&\hspace{30pt}+\mathcal{F}^{+}(\Omega_{s},{\bf
   k})+\mathcal{F}(\Omega_{s},{\bf k})\Big) \nonumber .
\end{eqnarray}
The normal Green's function $\mathcal{G'}$ and the anomalous
Green's functions $\mathcal{F}^{+}$ and $\mathcal{F}$ are defined
as
\begin{eqnarray}
   \mathcal{G'}(\tau,{\bf x}) &=& - \Big\langle
    T\: \psi'(\tau,{\bf x}) \psi'^{+}(0,0) \Big \rangle, \nonumber\\
   \mathcal{F}^{+}(\tau,{\bf x}) &=& -\frac{1}{n_{0}}
   \Big\langle  T\:\xi\: \xi \: \psi'^{+}(\tau,{\bf x})
   \psi'^{+}(0,0) \Big \rangle,  \label{greensfunctions}\\
   \mathcal{F}(\tau,{\bf x}) &=&-\frac{1}{n_{0}}
   \Big\langle T\:\xi^{+}\: \xi^{+} \: \psi'(\tau,{\bf x})
   \psi'(0,0) \Big \rangle. \nonumber
\end{eqnarray}
Note, that $\mathcal{G}'$ is related to the depletion of the
condensate, $n_{\rs D}= \mathcal{G'}(\tau\rightarrow 0^{+})$.
Within first order in the expansion parameter $\sqrt{n a^{3}}$,
the Green's functions take the Bogoliubov form
\cite{hugenholtz59,abrikosovbook}
\begin{eqnarray}
\mathcal{G'}(\Omega_{s},{\bf k}) &=& \frac{i \Omega_{s}+
\epsilon_{\bf k}+ \mu}{(i\:\Omega_{s})^2-E^{2}_{\bf k}}, \\
\mathcal{F}^{+}(\Omega_{s},{\bf k})&=& \mathcal{F}(\Omega_{s},{\bf
k})= -\frac{ \mu}{(i\:\Omega_{s})^2-E^{2}_{\bf k}},
\end{eqnarray}
with the excitation spectrum $E_{\bf k}= \sqrt{\epsilon^{2}_{\bf
k}+2 \epsilon_{\bf k} \:\mu }$ and the chemical potential $\mu = g
n$. The analytic continuation of $D_{0}(\Omega_{s},{\bf k})$
provides the contribution
\begin{equation}
  \chi_{0}(\omega,{\bf k})=
  \frac{2 \epsilon_{\bf k} \:  n_{0}}{\left(\hbar \omega+i\delta\right)^2-E^{2}_{\bf
  k}} \label{chi0}
\end{equation}
to the response. The structure factor  $S(\omega,{\bf k})$ is
related to the imaginary part of the response function, ${\rm Im}
\chi=- \pi [1-\exp(- \beta \hbar \omega)]S$, and using
(\ref{chi0}) we find ($\beta = 1 / T$)
\begin{equation}
  S_{0}(\omega,{\bf k})\!=\!\frac{n_{0}\; e^{\beta \hbar\omega}}{e^{\beta
  \hbar\omega}\!-\!1}\:
  \frac{\epsilon_{\bf k}}{E_{\bf k}}\:
  \Big[\delta\left(\hbar\omega \!- \! E_{\bf k}\right)
  +\delta\left(\hbar\omega \!+ \!E_{\bf
  k}\right)\Big],
\label{structur0}
\end{equation}
accounting for the excitation of single quasi-particles from the
condensate, the process illustrated in
Fig.~\ref{dissipativeprocesses}(a). Furthermore,
$S_{0}(\omega,{\bf k})$ nearly exhausts the $f$-sum rule
\begin{equation}
 \int_{-\infty}^{\infty} d\omega \: \omega  \: S_{0}(\omega,{\bf q}) =
 \frac{n_{0} \hbar^{2} {\bf q}^{2}}{2 m}; \label{fsumrule}
\end{equation}
the missing part  $n_{\rs D}\hbar^{2}{\bf q}^{2}/2m$ derives from
the structure of the thermal and quantum depletion of the
condensate. With $n_{\rs D}\propto  n\sqrt{na^{3}}$, its effect is
usually ignored to lowest order in the expansion $\sqrt{na^{3}}$
\cite{zambelli00,stamper-kurn99,steinhauer02}, but with increasing
interaction its contribution plays a substantial role.

The first diagram in the correction $D'=D'_{\rs(1)}+D'_{\rs (2)}$,
see Fig.~\ref{Dcombineddiagram}, takes the form
\begin{displaymath}
     D'_{\rs (1)}(\Omega_{s},{\bf k})\!= \!-T\! \sum_{t\in {\bf Z}} \!\int \!\!d{\bf q} \frac{i
     \Omega_{t}\!+\!
     \epsilon_{\bf q}\!+\!\mu}{(i \Omega_{t})^{2}-E^{2}_{\bf
    q}} \; \frac{i \Omega_{s+t}\!+\! \epsilon_{{\bf k}\!+\!{\bf
    q}}
    +\mu}{(i\Omega_{s+t})^{2}-E^{2}_{{\bf k}+{\bf q}}} .
\end{displaymath}
The term $\propto \mu^{2}$ diverges logarithmically for ${\bf k}
\rightarrow 0$; this divergence is removed in the full
perturbation theory \cite{hugenholtz59,castellani97} and results
in a higher order contribution of order $n a^{3} \ln(n a^3)$.
Similarly, the second diagram $D'_{\rs (2)}$ in
Fig.~\ref{Dcombineddiagram} contributes a correction of order $n
a^{3} \ln(n a^3)$. In the following analysis, we drop these terms
and account for corrections of order $\sqrt{n a^{3}}$.
Summation over Masubara frequencies $\Omega_{t}$ and analytic
continuation provides us with
\begin{widetext}
\begin{eqnarray}
 \chi'(\omega,{\bf k})&=& \int \frac{d{\bf q}}{(2\pi)^{3}} \frac{1}{2 E_{\bf q}}
 \left\{ \frac{E_{\bf q}+\epsilon_{\bf q}+ \mu}{e^{\:\beta E_{\bf
 q}}-1} \left[\frac{\hbar \omega + E_{\bf q}+ \epsilon_{{\bf q}+{\bf k}}+
 \mu}{\left[\hbar \omega+i \delta+E_{\bf q}\right]^{2}-E_{{\bf q}+{\bf k}}^{2}}
 +\frac{-\hbar \omega + E_{\bf q}+ \epsilon_{{\bf q}-{\bf k}}+
\mu}{\left[ \hbar \omega+i\delta-E_{\bf q}\right]^{2}-E_{{\bf
q}-{\bf
k}}^{2}}\right]\right. \label{chiprime}\\
&& \hspace{65pt}\left. + \frac{-E_{\bf q}+\epsilon_{\bf q}+
\mu}{1-e^{-\beta E_{\bf
 q}}} \left[\frac{\hbar \omega - E_{\bf q}+ \epsilon_{{\bf q}+{\bf k}}+
 \mu}{\left[\hbar \omega+i \delta-E_{\bf q}\right]^{2}-E_{{\bf q}+{\bf k}}^{2}}
 +\frac{-\hbar \omega - E_{\bf q}+ \epsilon_{{\bf q}-{\bf k}}+
\mu}{\left[\hbar \omega+i\delta+E_{\bf q}\right]^{2}-E_{{\bf
q}-{\bf k}}^{2}}\right] \right\} .\nonumber
\end{eqnarray}
\end{widetext}
The poles account for  the creation of two particles out of the
condensate with energy $\hbar \omega =E_{{\bf k}}+E_{{\bf k+q}}$
[see Fig.~\ref{dissipativeprocesses}(b)] and the scattering of
thermally excited quasi-particles $\hbar \omega =E_{{\bf
k}}-E_{{\bf k+q}}$ [see Fig.~\ref{dissipativeprocesses}(c)]. This
creation of particle hole-pairs dominates in the high temperature
limit $T \gg \mu$. Then, the response derives from
Eq.~(\ref{chiprime}) taking the limit $\mu \rightarrow 0$ and
reduces to the bosonic Lindhard function
\begin{equation}
 \chi'(\omega,{\bf k})=\int \frac{d{\bf q}}{(2\pi)^{3}}
 \frac{f\left[\epsilon_{\bf q}\right]-f\left[\epsilon_{{\bf q}
 +{\bf k}}\right]}{\hbar \omega +\epsilon_{\bf q}-\epsilon_{{\bf
 q}+{\bf k}}+i \delta}
 \label{finitetemperatureresponse}
\end{equation}
with $f\!=\!1/[\exp( \beta \epsilon)\! -\!1]$ the bosonic
distribution function.

\begin{figure}[hbtp]
\vspace{-0.cm}
\includegraphics[scale=0.4]{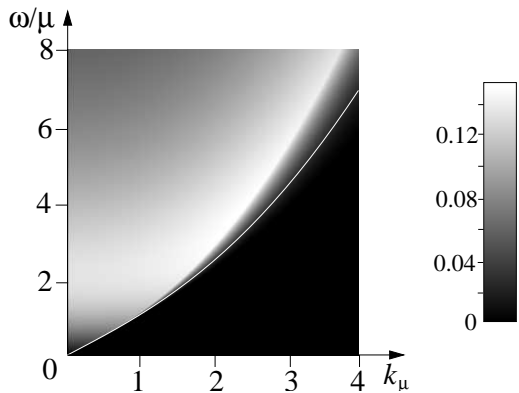}
\vspace{-0.2cm}
  \caption{ \label{structurefactor} Density plot
  of the dimensionless structure factor $\hbar \omega \tilde{S}'/2 \mu$
  (\ref{structureprime}) for an interacting gas at zero temperature;
  the momentum $k_{\mu}$ is measured in units  $\sqrt{ m \mu}/\hbar$.
  The white line denotes the energy threshold for the two-particle
  creation. At low momenta $|{\bf k}|< 2 \sqrt{ m \mu}/\hbar$, the peak
  in the structure factor appears at $\hbar \omega \approx 2 \mu$, while at
  high momenta  $|{\bf k}|>2 \sqrt{ m \mu}/\hbar$ this peak
  follows the Bogoliubov energy $\hbar \omega = E_{\bf k}$. 
  }
\end{figure}

Here, we are interested in the low temperature limit $T\ll \mu$ of
the response, where thermally excited quasi-particles are quenched
and the response function (\ref{chiprime}) accounts for the
quantum depletion alone. We set $T=0$, and
obtain the correction to the
dynamic structure factor
\begin{equation}
  S'(\omega,{\bf k}) =  n_{\rs D}\;
  \frac{\hbar^{2} {\bf k}^{2}}{2 m} \; \frac{1}{4 \mu^{2}}
  \; \tilde{S}'\left(\frac{\epsilon_{{\bf k}}}{4 \mu},\frac{\hbar \omega}{2
  \mu}\right)\label{structureprime}
\end{equation}
with $n_{\rs D}= (8/3 \sqrt{\pi}) \: n\:\sqrt{na^{3}}$ the quantum
depletion within Bogoliubov theory \cite{abrikosovbook}.
The dimensionless function $\tilde{S}'\left(\alpha,x\right)$
vanishes below the two-particle excitation threshold $x^{2}<
\alpha^{2}+ 2 \alpha$, while for $x^{2}> \alpha^{2}+ 2 \alpha$ we
obtain
\begin{displaymath}
  \tilde{S}'\left(\alpha,x\right) = \frac{3}{8 \sqrt{2}\:\alpha^{3/2}}
  \int_{\varepsilon^{-}}^{\varepsilon^{+}} d
  \varepsilon \left[ 1-
  \frac{1}{\left(\alpha+1+\varepsilon\right)^{2}-x^{2}}\right]^{1/2}\!\!\!,
\end{displaymath}
where  $\varepsilon^{-}=-1-\alpha+\sqrt{1+x^{2}}$. The upper
integration limit $\varepsilon^{+}$ is determined by
$\varepsilon_{1}< \varepsilon^{+}=\varepsilon_{2}<\varepsilon_{3}$
with $\varepsilon_{1,2,3}$ the solutions of the cubic equation
\begin{displaymath}
x^{4} - \left[\left(\alpha + \varepsilon +1\right)^{2} -1 + 4
\alpha \varepsilon\right] x^{2} + 4 \alpha \varepsilon
\left(\alpha + \varepsilon + 1\right)^{2} = 0.
\end{displaymath}
Combining the results (\ref{structur0}) and (\ref{structureprime})
provides the dynamic structure factor $S(\omega,{\bf k})=
S_{0}(\omega,{\bf k})+S'(\omega,{\bf k})$ within first order in
the expansion parameter $\sqrt{na^{3}}$. Using the relation
$\int_{0}^{\infty}dx \: x \tilde{S}'(\alpha,x) = 1$, the $f$-sum
rule is satisfied as required in  a consistent calculation to
order $\sqrt{na^{3}}$.
The dimensionless function $\hbar \omega \tilde{S}'/2\mu$ is shown
in Fig.~\ref{structurefactor}. At low momenta $|k|<\sqrt{ m
\mu}/\hbar$, the main contribution appears at high energies $\hbar
\omega \approx 2 \mu$ and accounts for the excitation of two
quasi-particles with opposite momenta, see
Fig.~\ref{dissipativeprocesses}(b). This  peak at low momenta is a
characteristic of the quantum depletion in an interacting Bose
gas.

\begin{figure}[htbp]
\vspace{-0.cm}
\includegraphics[scale=0.4]{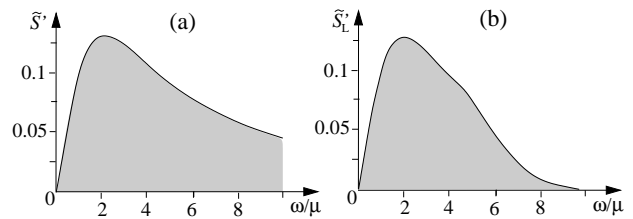}
\vspace{-0.2cm}
  \caption{ \label{comparison}
  Structure factor $\hbar\omega \tilde{S}'/2 \mu$ (a) and $\hbar\omega \tilde{S}'_{\rs L}/2
  \mu$ (b) in the limit of small momenta ${\bf k} \rightarrow 0$. The structure
  factor exhibits a peak at $\omega \approx 2 \mu$, and is quenched above twice the
  bandwidth $\approx 24 J$ in the presence of an optical lattice (b). }
\end{figure}

Subjecting the Bose gas to an optical lattice, we have to replace
the free dispersion relation $\epsilon_{\bf k}$ by the band
structure of a particle in a periodic potential (in the tight
binding limit, $\epsilon_{\bf k} \!=\! 2 J \left[3\!-\! \cos(k_{x}
b)\!-\!\cos(k_{y}b)\!-\! \cos(k_{z}b) \right]$ with the hopping
energy $J$ and the lattice constant $b$). In addition, Umklapp
processes transferring momentum to the optical lattice have to be
considered. Restricting the analysis to the lowest Bloch band and
small momenta ${\bf k} \rightarrow 0$, the expressions for the
response function (\ref{chi0}) and (\ref{chiprime}) remain valid
and the contribution $S_{\rs L}'(\omega,{\bf k})$ to the structure
factor takes the form
\begin{equation}
  S_{\rs L}'(\omega,{\bf k})
  = \frac{2 \mu^{2}}{(\hbar\omega)^{4}} \int_{K} \frac{ d{\bf
  q}}{(2 \pi)^{3}}
  \left({\bf k} \cdot \nabla \epsilon_{{\bf q}}\right)^{2}
  \delta(\hbar \omega- 2 E_{{\bf q}}) \label{periodicstructurefactor}
\end{equation}
with $K$ the first Brioulline zone. Furthermore, the structure
factor turns quasi-periodic, $S'_{\rs L}(\omega,{\bf k}+{\bf
K})=|c_{\bf\scriptscriptstyle K}|^2 S'_{\rs L}(\omega,{\bf k})$
for ${\bf k} \rightarrow 0$ with ${\bf K}$ a reciprocal lattice
vector and $c_{\bf\scriptscriptstyle K}$ a numerical prefactor
accounting for the  shape of the wave function within a potential
well (c.f., Ref.~\cite{menotti03} for a similar discussion of the
quasi-periodicity). In Fig.~\ref{comparison}(b) we show the
structure factor $ \tilde{S}'_{\rs L}(\hbar \omega/2 \mu) = 8
m^{*} \mu^{2} S'_{\rs L}(\omega,{\bf k})/(n_{\rs D} \hbar^{2} {\bf
k}^{2})$ in the tight binding- and ${\bf k} \rightarrow 0$ limit
with the effective mass $m^{*} = \hbar^{2}/(2 J b^{2})$.

In a recent experiment, St\"oferle {\it et al.} \cite{stoferle03}
have determined the energy transfer to an atomic gas via a
modulation of the optical lattice with frequency $\omega$. This
modulation perturbs the Bose gas with an external periodic
potential of amplitude $\delta V(\omega, {\bf K})$ and ${\bf K}$ a
reciprocal lattice vector. The energy per particle $\overline{W}$
transferred to the Bose gas during the time $\Delta t$ takes the
form \cite{chaikin} ($T=0$)
\begin{equation}
 \overline{W}= \frac{\pi}{2 } \:\frac{\Delta t}{n}\:
  \delta V^{2}(\omega, {\bf K})
  \:\hbar \omega \;\widehat{S}(\omega,{\bf K}).
  \label{power}
\end{equation}
The main energy transfer observed in \cite{stoferle03} resides at
large energies and is incompatible with the vanishing spectral
weight in $S_0$ at these frequencies \cite{menotti03}; in the
following we attempt an explanation involving the quantum
depletion. The structure factor $\widehat{S}(\omega,{\bf K})$ has
to account for the finite trapping potential: the wave functions
describing the Bogoliubov excitations involve momenta with a
spread $\Delta {\bf k}$. The structure factor in a finite trap
derives from its counterpart in the homogeneous system via the
average $\widehat{S}(\omega,{\bf K})\approx \langle S'(\omega,{\bf
K}+{\bf q})\rangle_{|{\bf q}|<|\Delta {\bf k}|}$. In the
interesting regime $\hbar \omega \approx \mu$, the structure
factor $S'$ involves two-particle excitations with large momenta
and the finite level spacing in the trap plays a minor role. In
estimating the relevant spread $\Delta {\bf k}$ we have to account
for the trap size $R$ (Thomas Fermi radius) and the condensate
healing length $\xi$ describing the rapid decay of the wave
function at the trap boundary; making use of the excitations
derived in Ref.~\cite{stringari96}, we obtain a spread of order
$|\Delta {\bf k}|\approx 1/(R \xi)^{1/2}$. Combining the
quasi-periodicity of the structure factor in an optical lattice
and the spread induced by the trapping potential, we find the
estimate
\begin{equation}
   \widehat{S}(\omega,{\bf K})
   \approx |c_{\bf\scriptscriptstyle K}|^2 \langle S'_{\rs
   L}(\omega,{\bf q} )\rangle_{|{\bf q}|<|\Delta {\bf k}|}
   \label{Strap}
\end{equation}
with $S'_{\rs L}(\omega,{\bf k})$ shown in
Fig.~\ref{comparison}(b). In their experiments, St{\"o}ferle {\it
et al.} \cite{stoferle03} report (among other) on the energy
transfer to a trapped Bose gas subject to a 3D optical lattice
residing in the superfluid phase. They find a pronounced peak in
the excitation spectrum centered at a frequency $\hbar\omega
\approx 2\mu$ and with a width of order of the band width, in
rough agreement with our result for $S'_{\rs L}$. Accounting for
the smearing in the structure factor due to the finite trap as
discussed above, the transferred energy (\ref{power}) assumes the
correct order of magnitude. We then are tempted to associate the
observations made in \cite{stoferle03} with the quantum depletion
of the Bose gas enhanced by the presence of the optical lattice.
We note that the optical lattice applied in the experiments is
already appreciable. Hence, we expect the weak coupling analysis
above to reproduce correctly the qualitative features in the
experiment, while a quantitative discussion requires to include
higher order corrections in the parameter $\sqrt{n a^{3}}$.

In conclusion, the structure factor of a weakly interacting Bose
gas exhibits a characteristic behavior associated with
the structure of the quantum depletion of the condensate. A
measurement of the structure factor not only allows for the
detection of the quantum depletion in atomic gases but provides
additional non-trivial information on the ground state of
strongly interacting Bose gases, thus opening up new possibilities
in studying complex quantum phases in atomic gases.

 {\acknowledgments We thank T.\ Esslinger, M.\ K\"ohl, H.~G.\ Katzgraber, S.~Huber, and
W.\ Zwerger for stimulating discussions.}


\end{document}